\documentclass[%
 reprint,
 superscriptaddress,
 amsmath,amssymb,
 aps,
]{revtex4-2}

\usepackage{subfigure}
\usepackage{amsmath,graphicx,amssymb}
\usepackage{multirow,color}
\usepackage{url}
\usepackage{verbatim}
\usepackage{xcolor}

\usepackage{bm}

\newcommand{\beq}{\begin{equation}}
\newcommand{\eeq}{\end{equation}}

\begin{document}
\preprint{APS/123-QED}

\title{Euler-Maruyama method for non-Wiener processes}
\author{Richard D.J.G. Ho}
\affiliation{Njord centre, Department of Physics, University of Oslo, Norway}

\date{\today}

\begin{abstract}
Descriptions of complex physical or biological systems often include stochastic contributions, and these are commonly simulated using Wiener processes.
In many cases however, non-Gaussian fluctuations may originate from non-Wiener processes which remain less explored.
The Euler-Maruyama method of discretising stochastic differential equations to non-Wiener processes is generalised.
Non-Gaussian noise generated from a subset of L\'evy processes can be used simply and often with more physical justification, for both additive and multiplicative noise.
An example of this is provided that gives superior physical results compared to using geometric Brownian motion.
Finally the results of the additive noise are shown to be equivalent to a derived master equation via the Kramers-Moyal expansion.
\end{abstract}

\maketitle

\section{Introduction}

Non-Gaussian fluctuations are frequently encountered in non-equilibrium or with complex interactions. 
For instance, many biological systems have spatio-temporal dynamics marked by fluctuations which span across multiple scales, from large scale collective dynamics down to the cellular and sub-cellular levels \cite{elowitz2002stochastic}.
Overall, these different sources contribute to non-Gaussian fluctuations which help regulate biological functions such as transcription, decision making, morphogenesis, statistical ecology, stable pattern formation, and
biochemical reactions \cite{tsimring2014noise,eling2019challenges,majka2023stability,dennis1984gamma}.
As well, in many areas outside of biology and statistical physics, it is common to use processes with non-Gaussian distributions,
such as the simple gamma process used in degradation modeling \cite{park2005accelerated,rodriguez2018degradation},
and maintenance \cite{van2009survey}, 
or the variance-gamma process in financial modeling \cite{madan1990variance}, which is the difference of two gamma processes \cite{avramidis2006efficient}.

Models including noise allow a stochastic understanding of biochemical and other processes, and attempts have been made to simply and improve the procedure for implementing them \cite{manninen2006developing,browning2020identifiability,wilkinson2009stochastic,ditlevsen2013introduction}.
The simplest case is that of a Gaussian and Markovian (white) process.
Whilst straightforward techniques exist to generate time correlated (colored) Gaussian noises \cite{romero1999generation}, it is also possible to introduce time correlated non-Gaussian noises using a subordinate process.
Here, the potential of a stochastic sub-process is often implemented with a carefully chosen potential to ensure that the stationary distribution of the sub-process is of a desired form.

Subordinate implemented noises will necessarily have their own timescales, and this can interact non-trivially with the time scale of the main process of interest.
To avoid this problem, in this paper the Euler-Maruyama method is extended to a wider range of processes than the Wiener process (which uses a Gaussian distribution).
It is further shown how this can be done with weakly non-linear functions of a normally distributed subordinate noise.
These methods are applicable to both additive and multiplicative noise,
the latter of which has not been seen in the literature to the best of our knowledge.
We finally recapitulate the dynamics of a Kramers-Moyal expansion to show how the dynamics of the problem can be captured with a master equation.

\section{Relaxed Euler-Maruyama Method}

Consider the problem of modeling the evolution of some noisy process numerically. When the timescale of the noise, $\tau_n$, is smaller than the simulation time step, $\Delta_t$, we may resort to the Euler-Maruyama method for the associated variable $x$
\begin{align}
    x(t + \Delta_t) = a(x,t) \Delta_t + b(x,t) L (\Delta_t) \ ,
    \label{eq:EM}
\end{align}
where we use random increments $L(\Delta_t)$.
For the strict Euler-Maruyama method, we take $L (\Delta_t) \sim \mathcal{N}(0,\sigma^2 \Delta_t) \sim \sigma \sqrt{\Delta_t} \mathcal{N} (0, 1)$.
For this to give consistent results, it is not a requirement that $\tau_n \rightarrow 0$, merely that $\tau_n \ll \Delta_t$.
For the following, 
no claim is made about application to strictly white noise,
instead only noise to where the correlation time $\tau_n$ is very small compared to the desired simulation time step and system timescale.

In this paper, we make the claim that,
for suitable functions $b(x,t)$,
we can relax the Euler-Maruyama requirement
and allow the increment to be drawn from any distribution $\mathcal{L}(\Delta_t)$, which has an associated random variable $L(\Delta_t)$ satisfying
\begin{align}
\langle L \rangle &\sim \mathcal{O}(\Delta_t)  \nonumber \\
\langle L^2 \rangle - \langle L \rangle^2 &\sim \mathcal{O}(\Delta_t) \nonumber \\ 
\sum_n L(\delta_t) &\sim L(\Delta_t) \ , \ n \delta_t = \Delta_t  \nonumber \ ,
\end{align}
where angled brackets represent an average, and $\sim$ statistical similarity.
The first two properties ensure that our distribution is well behaved and specifically excludes many members of the stable distribution.
The last property is a restricted form of infinite divisibility.
It states that the sum of independent random variables, called their convolution, follows the same type of distribution as the random variables themselves.
The gamma distribution, the normal distribution, as well as many other distributions, have this property.
The random variable will then follow
\begin{align}
L(t) = \int_0^t dL_s \sim \mathcal{L}(t) \ , \nonumber
\end{align}
and this will hold for all higher order moments, such as skewness and kurtosis, by definition.

\if
In mathematical terms, the above is an example of a L\'evy process \cite{bertoin1996levy,applebaum2009levy},
which should not be confused with the L\'evy distribution.
Much work has been done in proving the convergence of the Euler-Maruyama method 
for much more general L\'evy processes
than the ones described here, such as in \cite{kuhn2019strong}, which proves it for the case $b = 1$.
However, since our process is restricted, but may be useful physically,
we can use a simple adaptation of a proof for more general forms of $a$ and $b$ 
such as that of Theorem 2.2 in 
\cite{higham2002strong},
by noting that our restricted L\'evy process is a martingale.
\fi

An intuitive understanding for why this method works is to consider the following about using a subordinate process to simulate a noisy process, $\eta(t)$, with associated timescale $\tau_n$.
Here, the evolution is modeled as
\begin{align}
    x(t + \Delta_t) &= a(x, t) \Delta_t + b(x, t) f(\eta(t)) \Delta_t \ ,
\end{align}
where $f(\eta(t))$ is some function
and $\eta(t)$ is generated using an algorithm such as that found in \cite{romero1999generation}.
The Euler method can be used provided that $\Delta_t \ll \tau_n, \tau_s$, where $\tau_s$ is the timescale of this system without noise.

Now, it should be noted that the following distributions are identical
\begin{align}
    \int_0^t f(\eta(t^*)) \Delta_t dt^* \sim \sum_n L (\Delta_t) \ . \nonumber
\end{align}
if $\mathcal{L}(\Delta_t)$ is chosen with the properties listed here.
Since we are not dealing explicitly with a white noise, the ``power'' contributions to the moments of $x$ will be non-infinite. This allows use of different distributions than just the normal distribution.
By relaxing the Euler-Maruyama method, the interaction between the deterministic parts of the evolution and the tails of the distribution can be captured, which may be important. This allows us to replace $f(\eta(t)) \Delta_t$ of the subordinate method with $L(\Delta_t)$ in this situation.

A natural problem is to consider $f(\eta(t))$ as a weakly non-linear function of a Gaussian distributed noisy process.
We can apply the relaxed Euler-Maruyama method provided that $\eta(t)$ has a small variance, $\sigma^2$, compared to the convergence of the Taylor expansion of $f(\eta)$.
We take the Taylor expansion
\begin{align}
f(\eta) = f(\mu) + f'(\mu) (\eta - \mu) + \frac{f''(\mu)}{2} (\eta - \mu)^2 + \ ... \nonumber
\end{align}
with $\mu$ being the mean of $\eta$
and substitute $\eta = \mu + X$, where $X \sim \mathcal{N} (0, \sigma^2 )$.
This gives
\begin{align}
F(\mu + X) = F(\mu) + F'(\mu) X + \frac{F''(\mu)}{2}X^2 +  \ ... \nonumber
\end{align}
However, $X$ and $X^2$ are not independent, so cannot be replaced with $\mathcal{N} ( 0, \Delta_t \sigma^2 )$ or $\sigma^2 \frac{F''(\mu)}{2} \chi^2(\Delta_t)$ respectively, unless $F'(\mu) = 0$ or $F''(\mu) = 0$.
Otherwise, we complete the square,
\begin{align}
F(\mu + X) = F(\mu) + F'(\mu) \sigma \mathcal{N}(0,1)  + \frac{F''(\mu)}{2} \sigma^2 \mathcal{N}(0,1)^2 + \ ... \nonumber \\
= F(\mu) - \frac{F'(\mu)^2}{2 F''(\mu)} + \frac{F''(\mu)}{2} \sigma^2 \mathcal{N} \bigg(\frac{F'(\mu)}{F''(\mu) \sigma},1 \bigg)^2 +  \ ... \nonumber
\end{align}
This last term follows a non-central $\chi^2$ distribution \cite{patnaik1949non}, $\chi'^2(k, \lambda)$, where $k$ is the number of degrees of freedom and $\lambda$ the square of the mean. Here we have $k = 1$. For zero mean, $\chi'^2(k, 0)$, we recover the usual $\chi^2 (k)$ distribution.
The non-central $\chi^2$ distribution follows
\begin{align}
\sum_i^n \chi'^2(k_i, \lambda_i) &\sim \chi'^2 \bigg( \sum_i^n k_i, \sum_i^n \lambda_i \bigg) \ , \nonumber 
\end{align}
which gives the small expansion approximation for the Euler-Maruyama method
\begin{align}
L (\Delta_t ) \sim \ &\Delta_t \bigg( F(\mu) - \frac{F'(\mu)^2}{2 F''(\mu)} \bigg) \nonumber \\
&+ \ \frac{F''(\mu)}{2} \sigma^2 \chi'^2 \bigg(\Delta_t , \Delta_t \frac{F'(\mu)^2}{F''(\mu)^2 \sigma^2} \bigg) \ .
\label{eq:smallexpansion}
\end{align}
This can be inserted into Eq.~(\ref{eq:EM}).

\section{Comparison of methods for additive noise}
\label{sec:additive}



We use the Euler-Maruyama method as in Eq.~(\ref{eq:EM}) to model a noisy exponential decay with additive noise,
where $a = -\nu X_t$ and $b = 1$ for different L\'evy processes,
\begin{align}
dX_t = -\nu X_t dt + dL_t \ .
\label{eq:rde}
\end{align}
We make two comparisons.
The first is to what we call the naive method, which uses $\Delta L_n \sim \Delta_t \mathcal{L}(1)$.
Using the property of infinite divisibility for the L\'evy process shows that the Euler-Maruyama method
is equivalent to the naive method for $\Delta_t = 1$.
This makes it the natural continuous extension for processes originally modeled with fixed unit time step,
as in going from a random walk to Brownian motion.
The second comparison is to an immediate invocation of the central limit theorem.
Here, we replace the underlying distribution by an equivalent normal distribution with mean and variance equal to those used in the relaxed Euler-Maruyama method.
The different increments and distributions are summarised in Table.~(\ref{tbl:params1}).

\begin{table*}
\caption{Increments, $\Delta L_n$, used in simulations of Eq.(\ref{eq:rde}) with results shown in Fig.~\ref{fig:rd}.
  For the Euler-Maruyama method we approximate C with Eq.~(\ref{eq:smallexpansion}).}
  \label{tbl:params1}
\begin{ruledtabular}
\begin{tabular}{cccc}

process & naive & Euler-Maruyama & central limit theorem \\
    \hline
    (a) & $\Delta_t (\Gamma (1,1) - \Gamma (1,1))$ & $\Gamma (\Delta_t,1) - \Gamma (\Delta_t,1)$ & $\mathcal{N}(0,2 \Delta_t)$  \\
    (b) & $\Delta_t \Gamma (2,1/2)$ & $\Gamma (2 \Delta_t,1/2)$ & $\mathcal{N}(\Delta_t,\Delta_t/2)$  \\
    (c) & $\Delta_t \cos(\mathcal{N}(0,1/4))$ & $\cos(X) \ , \ X \sim \mathcal{N}(0,1/4)$ & $\mathcal{N}(7\Delta_t/8,\Delta_t/64)$
\end{tabular}
\end{ruledtabular}
\end{table*}

\begin{figure}
\centering
\includegraphics[width=0.43\textwidth]{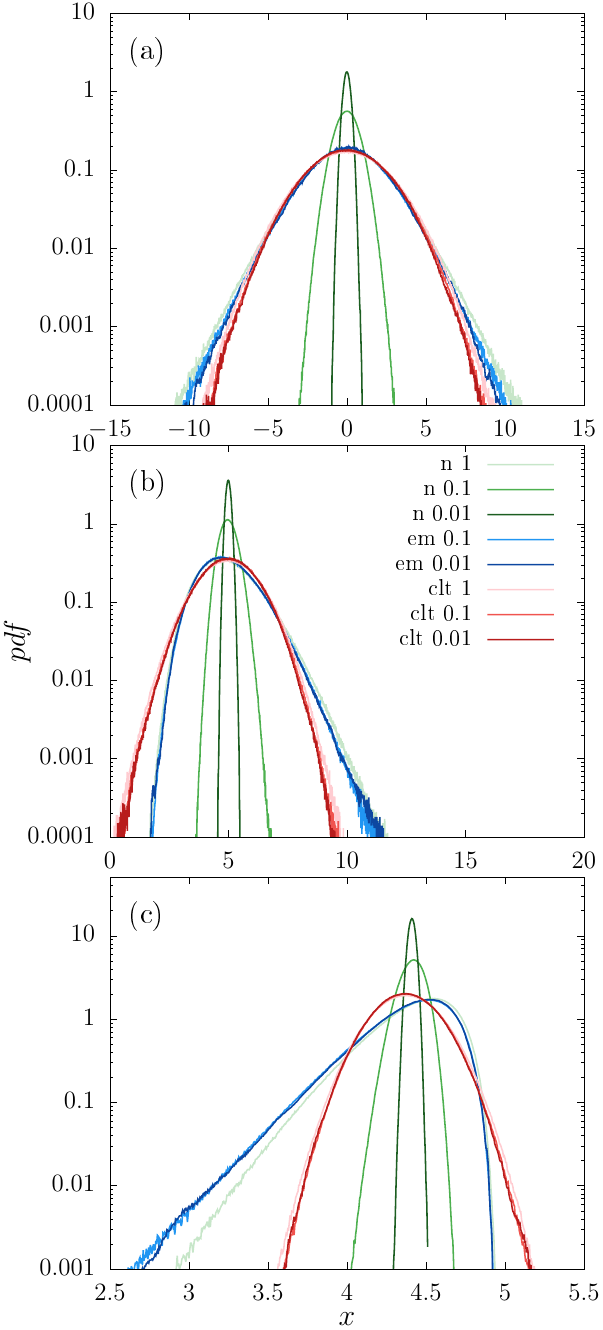}
\caption{Probability density functions of the reaction-diffusion system for different L\'evy processes,
comparing the naive ($n$), Euler-Maruyama ($em$), and central limit theorem ($clt$) methods. The letter refers to Table.~(\ref{tbl:params1}), whilst the
number refers to the time step length.}
\label{fig:rd}
\end{figure}

Process ($a$) is the difference of two gamma distributions, process ($b$) is a single gamma distribution, and process ($c$) is the case where $f(\eta(t)) = \cos(\eta(t))$. In process ($c$), for the relaxed Euler-Maruyama method we approximate our process by a $\chi^2$ distribution.
For $\Delta_t = 1$, in process ($a$), $\Delta L_n$ is drawn from a Laplace distribution, which is a special case of
the variance-gamma distribution.
Simulations are performed with $\nu = 0.2$ and taking the probability densities generated by evolving the system for $10^7$ time steps.
The comparisons of the different methods are shown in Fig.~\ref{fig:rd}.
The distribution with exponential tails shown in Fig.~\ref{fig:rd} (a) is similar to that found in cell center displacements when simulating glassy dynamics of tissue fluidity \cite{ray2024role}.
The fact that it can be directly simulated here may prove useful to modelers of biophysical systems.

As can be seen in Fig.~\ref{fig:rd},
the naive approximation preserves the mean but results in decreasing variance with decreasing time step.
The probability density becomes more like a normal distribution when the time step is reduced; this is because the central limit theorem becomes important here.
However, whilst a direct application of the central limit theorem preserves the mean and variance of the resulting distribution,
it also obliterates all information of the longer tails of the distributions and any skewness, which is especially obvious in process ($b$).
For process ($c$), using cos$(X)$, the probability density has a slightly fatter tail in the negative direction.
This occurs because the expansion is truncated at $X^2$.
The next term in cos$(x)$, $X^4$, would have the tendency to push the system toward higher $x$.

\section{Comparison of methods for multiplicative noise}

We now show a comparison for modeling a noisy exponential decay with multiplicative noise.
A number of particles $N$, which decay randomly at rate $\lambda$, is modeled here in two ways.
First, we use a standard Weiner process ($w$) and generate a form of geometric Brownian motion, with
\begin{align}
N(t + \Delta_t) &= N(t) - \lambda N(t)(\Delta_t + \Delta W_n) \nonumber \\
&= N(t) - N(t) \Delta W'_n \nonumber \\
\Delta W_n &\sim \mathcal{N}(0,\Delta_t) \ , \ \Delta W'_n \sim \mathcal{N}(\lambda \Delta_t,\lambda^2 \Delta_t) \ . \nonumber
\end{align}
In the second, we use the relaxed Euler-Maruyama method for a gamma process ($g$)
\begin{align}
N(t + \Delta_t) = N(t) - N(t) \Delta L  \ , \ \Delta L \sim \Gamma(\Delta_t, \lambda) \ . \nonumber
\end{align}
The justification for the second equation comes from the exponential distribution.
The exponential distribution is the probability distribution of the time between events in a Poisson point process,
such as the decay we are modeling. In one time unit, the system on average experiences $1/\lambda$ decays.
We multiply that by the population to make the number of decays proportional to the population 
and then subtract that value from the original population. 
We appeal to infinite divisibility by noting
that the exponential distribution is the gamma distribution with shape parameter $k = 1$.

\begin{figure}
\centering
\includegraphics[width=0.47\textwidth]{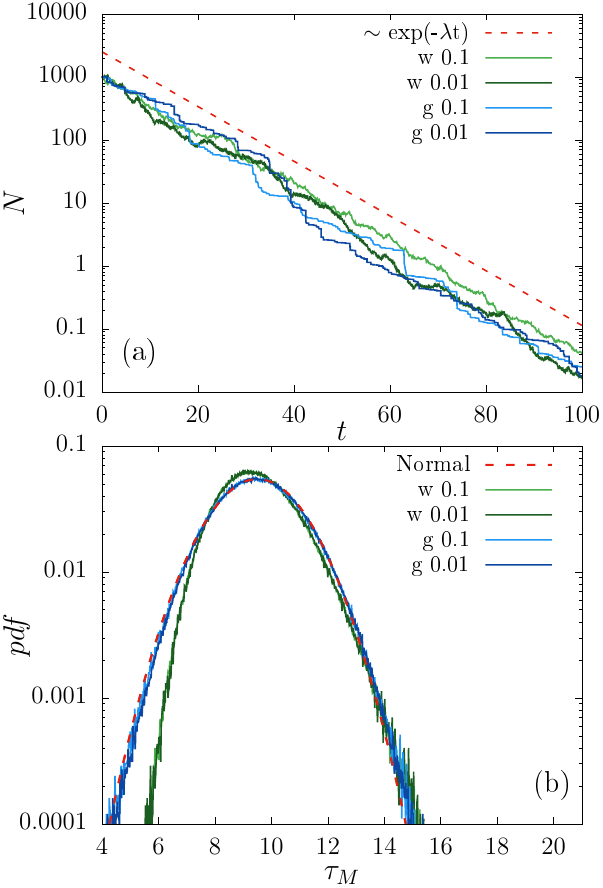}
\caption{Modeling of noisy exponential decay. a) Examples of decays for the standard Wiener process ($w$) and gamma process ($g$). b) The measured mean lifetime of the simulations. The number refers to the time step.}
\label{fig:ExpoDecay}
\end{figure}

Fig.~\ref{fig:ExpoDecay} shows the comparison between simulating the Weiner process ($w$) and gamma process ($g$)
for $\lambda = 0.1$ which implies lifetime $\tau = 10$.
The system was evolved until it reached $e^{-5}$ of its original value, which was $10^3$, 
and this time was divided by five to give the measured mean lifetime, $\tau_M$.
This measurement was performed $10^5$ times.
Assuming a normal distribution, for the Wiener process decay with time step 0.01, we find $\tau_M = 9.5 \pm 1.3$, whereas for the gamma process decay $\tau_M = 9.5 \pm 1.5$.
However, the Wiener process result is clearly skew, as can be seen in the figure.
Given that we are taking repeated results, we would expect, by the central limit theorem, that our measured value should approximate a normal distribution. This is the case when we have modeled the decay using a gamma process, and shows how it may capture important properties of derived statistics.

The non-zero variance in Fig.~\ref{fig:ExpoDecay} (b) is caused by the fact that we have evolved the system over five mean lifetimes,
if we evolved it longer, we would expect decreasing variance.
The reason the measured value is not exactly 10 is because of the weakness of how the Euler-Maruyama
method treats the mean values, not in how it treats noise.
For instance, if we use the classic Runge-Kutta method, but treat the L\'evy process
as fixed over the time step, we get a value of 10 \cite{hansen2006efficient}.
In this example we have shown how using the relaxed Euler-Maruyama method produces results that are more physical than when using a standard Weiner process and the strict Euler-Maruyama method.

\section{Master Equation for Relaxed Euler-Maruyama}

If we take a simplified relaxed Euler-Maruyama method $x(t + \Delta t) = x(t) + L(\Delta_t)$, where $L(\Delta_t)$ is drawn from some distribution, then
we can use the Chapman-Kolmogorov equation to derive
the evolution of the probability density
\begin{align}
    \partial_t P &\approx \sum_{n = 0}^\infty \frac{(-1)^n}{n!} \frac{\partial^n}{\partial x^n} (\langle L^*(\Delta_t)^n \rangle P) \ ,
    \label{eq:probevo}
\end{align}
with $P = P(x, t | x_0, t_0)$ and $\langle L^*(\Delta_t)^n \rangle$ being the term which is of order $\mathcal{O}(\Delta_t)$.
This is equivalent to the Kramers-Moyal expansion
\begin{align}
    \partial_t P(x, t) &\approx \sum_{n = 1}^\infty (-\partial_x)^n (D_n(x, t) P(x,t)) \ , \nonumber
\end{align}
but using different co-efficients.
For $L(\Delta t) \sim \sqrt{2D} \mathcal{N}(0, \Delta t)$ we recover the classic evolution for a normal distribution
\begin{align}
    \partial_t P = D \partial_{xx}P \ . \nonumber
\end{align}
However, with $L(\Delta t) \sim \Gamma(k \Delta t, \theta)$, we use the fact that the the raw moments for $\Gamma(k, \theta)$ go as $\theta^n \Gamma(k + n) / \Gamma(k)$, to show that, to $\mathcal{O}(\Delta_t)$, $\langle L(\Delta_t)^n \rangle = \theta^n (n - 1)!$.
This gives us a master equation for a pure gamma process of
\begin{align}
    \partial_t P = \sum_{n = 1}^\infty \frac{(-1)^n}{n} k \theta^n \frac{\partial^n P}{\partial x^n}  \ .
    \label{eq:gammaMaster}
\end{align}

\begin{figure}
\centering
\includegraphics[width=0.47\textwidth]{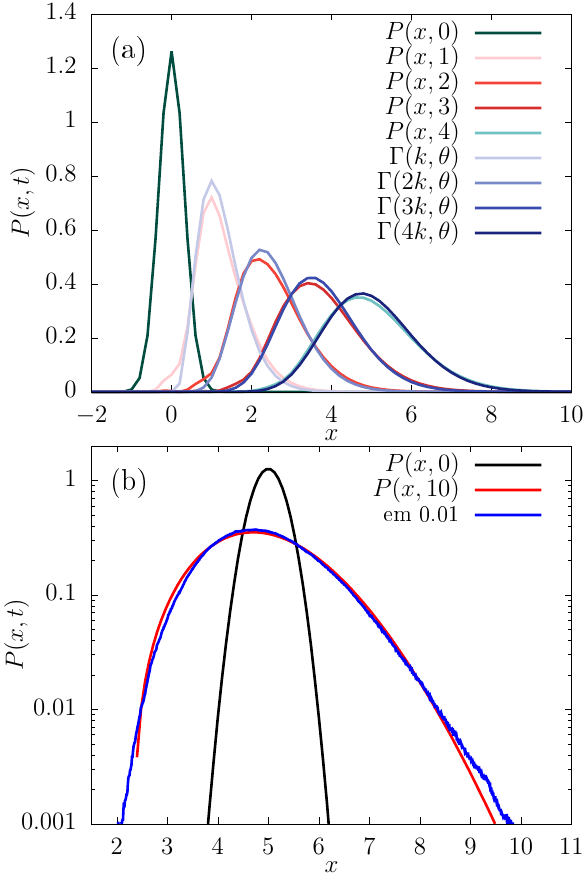}
\caption{a) Solution of equation Eq.~(\ref{eq:gammaMaster}) compared to the exact solution for an initial delta distribution at $x = 0$. b) for additive noise, compared to using EM method.}
\label{fig:KME}
\end{figure}

We now compare the prediction of Eq.~(\ref{eq:gammaMaster}) to directly using the 
relaxed Euler-Maruyama method.
The Pawula theorem states that we cannot truncate the series for an exact solution, but due to numerical limitations, this will be required.
We solve Eq.~(\ref{eq:gammaMaster}) with $k = 5, \theta = 1/4$ and initialise $P(x,0)$ with a Gaussian with $\sigma^2 = 0.1$ centered at zero. We take terms up to $n = 5$, then compare to the exact solution for the Euler-Maruyama method when we initialize with $\delta(0)$.
Due to the infinite divisibility of the gamma distribution, this is $\Gamma(kt, \theta)$.
The comparison is shown in Fig.~\ref{fig:KME} $(a)$ with good agreement.

For comparison with a different additive noise, we can use $L(\Delta_t) = -\nu x \Delta_t + \Gamma(k \Delta_t, \theta)$
to compare the evolution of Eq.~(\ref{eq:gammaMaster}) to process ($b$) in Section \ref{sec:additive}.
Fig.~\ref{fig:KME} (b) shows the comparison of simulations of Eq.~(\ref{eq:gammaMaster})
and of a long time average of evolution using Eq.~(\ref{eq:EM}).
Here we have $\nu = 0.2, \theta = 1/2, k = 2,$ and use orders up to $n = 2$, due to numerical instability.

\section{Discussion}

In this paper, we have shown how the classic Euler-Maruyama method can be relaxed 
by drawing the increments from specific non-Gaussian distributions, with suitable variation
of the distribution dependent on the time step.
In comparison to other methods of adding non-Gaussian noise via a subordinate noise,
this allows the noise to have a timescale that is smaller than the simulation time step.
We have further shown by looking at the random decay process an example where using this non-Gaussian step allows better representation of the half-life statistics.
Finally, we show that the dynamics can be reformulated into a master equation via the Kramers-Moyal expansion and show good agreement with our previous results using the Euler-Maruyama method.
We hope that these simplifications to using non-Gaussian noises will lower the barrier to entry for modelers in various fields, such
as biological physics,
where non-Gaussian noise is commonplace in reality but under-utilised in modeling due to its complexity.

\

\begin{acknowledgments}
The author benefited from insightful conversations with M. Majka, L. Angheluta and, J. K. Pierce.
\end{acknowledgments}

\bibliography{refs_EMnonGauss}

\end{document}